\documentclass[pre,aps,twocolumn,showpacs]{revtex4}
\usepackage{epsfig}
\usepackage{graphicx}
\begin{document}
\title{Giant fluctuations at a granular phase separation threshold}
\author{Baruch Meerson$^{1}$, Thorsten
P\"{o}schel$^2$,  Pavel V. Sasorov$^{3}$ and Thomas
Schwager$^{2}$}
 \affiliation{$^{1}$Racah Institute of Physics,
Hebrew University of Jerusalem, Jerusalem 91904, Israel}
\affiliation{$^2$Institut f\"ur Biochemie, Charit\'e, Monbijoustr.
2, 10117 Berlin, Germany} \affiliation{$^{3}$Institute of
Theoretical and Experimental Physics, Moscow 117218, Russia}
\begin{abstract}
We investigate a phase separation instability that occurs in a
system of nearly elastically colliding hard spheres driven by a
thermal wall. If the aspect ratio of the confining box exceeds a
threshold value, granular hydrostatics predict phase separation:
the formation of a high-density region coexisting with a
low-density region along the wall that is opposite to the thermal
wall. Event-driven molecular dynamic simulations confirm this
prediction. The theoretical bifurcation curve agrees with the
simulations quantitatively well below and well above the
threshold. However, in a wide region of aspect ratios around the
threshold, the system is dominated by fluctuations, and the
hydrostatic theory breaks down. Two possible scenarios of the
origin of the giant fluctuations are discussed.

\end{abstract}
 \maketitle
\section{Introduction} Dynamics of a system of inelastically
colliding hard spheres have attracted a great deal of recent
interest \cite{Kadanoff,GG}, in particular in the context of
validity of kinetic theory and hydrodynamics of rapid granular
flow developed in the 80-ies \cite{hydro}. Hydrodynamics looks
ideally suitable for a description of large-scale patterns
observed in rapid granular flows: a plethora of clustering
phenomena \cite{Kudrolli}, vortices \cite{vort}, oscillons
\cite{sw1}, shocks \cite{sw2}, etc., that are difficult to
understand in the language of individual particles. However, a
first-principle derivation of a universally applicable continuum
theory of granular gas is not a simple task, even in the dilute
limit. The use of the Enskog equation, the starting point of a
systematic derivation of the constitutive relations of granular
hydrodynamics, is based on the \textit{Molecular Chaos}
hypothesis. This hypothesis is justified for not too large
densities and for an ensemble of \textit{elastic} hard spheres.
Its use for \textit{inelastic} hard spheres is not obvious, as
inelasticity of the particle collisions introduces inter-particle
correlations \cite{correlations}. The correlations become stronger
as the inelasticity of the collisions increases. On the contrary,
for \textit{nearly elastic} collisions, $1-r^2 \ll 1$ (where $r$
is the coefficient of normal restitution) the correlations are
small, and the Enskog equation can be safely used.

An important additional assumption, made in the process of the
derivation of hydrodynamics from the Enskog equation, is scale
separation. Hydrodynamics demands that the mean free path of the
particles be much less than any characteristic length scale, and
the mean time between two consecutive collisions be much less than
any characteristic time scale described hydrodynamically. This
condition should be verified, in every specific system, after the
hydrodynamic problem is solved and the characteristic length and
time scales determined. Again, it is safe to say that this
condition can be satisfied if the particle collisions are nearly
elastic \cite{Esipov,Grossman,Goldhirsch}. Restrictive as it is,
the nearly elastic limit is conceptually important just because
granular hydrodynamics is expected to work here.

Another potentially important, albeit largely unexplored,
limitation of the validity of granular hydrodynamics (or, rather,
of any continuum approach to rapid granular flow) is due to the
noise caused by the discrete nature of particles. Noise is
stronger here than in classical (molecular) fluids simply because
the number of particles is much smaller. In addition, noise can be
amplified at thresholds of hydrodynamic instabilities as found,
for example, in Rayleigh-B\'{e}nard convection of classical fluids
\cite{Hohenberg}.

The validity of hydrodynamic description in general, and the
accuracy of constitutive relations in particular, can be
conveniently checked on symmetry-breaking instabilities that are
abundant in rapid granular flows.  The example of a
symmetry-breaking instability that we consider in this work deals
with a very simple setting: a two-dimensional (2D) system of
nearly elastically colliding hard spheres, confined by a
rectangular box and driven by a thermal sidewall at zero gravity.
The setting is described in detail in Sec. II. The basic steady
state here is the ``stripe state'': a stripe of enhanced density
at the wall opposite to the driving wall \cite{Grossman}. In the
continuum language, the stripe state is uniform in the lateral
direction, by which we mean the direction parallel to the driving
wall. Within a certain range of parameters (delineated below),
steady-state equations of granular hydrodynamics predict
spontaneous symmetry breaking instability of the stripe state,
when the aspect ratio of the confining box exceeds a certain
threshold \cite{LMS,Brey2,KM,LMS2}. The instability leads to phase
separation: the development of ``droplets'' (high-density domains)
coexisting with ``bubbles'' (low-density domains). For
\textit{very} large aspect ratios of the box, this
symmetry-breaking instability has been recently observed in
event-driven molecular dynamic (EMD) simulations, and described by
a phenomenological continuum model \cite{Argentina}. The present
work is devoted to a more detailed investigation of the phase
separation instability in the range of aspect ratios
\textit{comparable} to the threshold value. We employ, in Sec.
III,  the equations of granular hydrodynamics (or rather
hydrostatics) to compute the supercritical bifurcation curve for
the phase separation instability. Then we report, in Section IV,
on extensive EMD simulations that show that this bifurcation curve
is quantitatively accurate \textit{well} below and \textit{well}
above the threshold value of the aspect ratio. Unexpectedly, the
hydrostatic theory fails in a relatively \textit{wide} region of
aspect ratios around the threshold value, where the system is
found to exhibit giant fluctuations. In an attempt to get insight
into the mechanism of this anomaly, we investigate, also in
Section IV, the dependence of the magnitude of fluctuations on the
total number of particles in the system. A summary and discussion
of our results is presented in Section V.

\section{Model system and hydrostatic equations}

Let $N$ hard spheres of diameter $d$ and mass $m=1$ move in a 2D
rectangular box $L_x \times L_y$. The inelasticity of particle
collisions is parameterized by a constant coefficient of normal
restitution $r$. Particle collisions with three of the walls are
elastic. The fourth, thermal wall is located at $x=L_x$. Upon
collision with it, the normal component of the particle velocity
is drawn from a Maxwell distribution with temperature $T_0$
\cite{Grossman}, while the tangential component of the particle
velocity is preserved.

Working in the nearly elastic limit $1-r^2 \ll 1$ and employing
the Navier-Stokes hydrodynamics \cite{hydro}, we introduce the
number density $n(\mathbf{r},t)$, granular temperature
$T(\mathbf{r},t)$ and mean-flow velocity
$\mathbf{v}(\mathbf{r},t)$. Energy input at the thermal wall can
be balanced by the dissipation due to inter-particle collisions.
Therefore, we assume that the system reaches a zero-mean-flow
steady state $\mathbf{v}=\mathbf{0}$, and is therefore describable
by the simple momentum and energy balance equations:
\begin{equation}
p={\rm const}\,,\,\,\, \nabla \cdot (\kappa \, \nabla T) = I \,.
\label{energy1}
\end{equation}
Here $p$ is the pressure, $\kappa$ is the thermal conductivity and
$I$ is the rate of energy loss by collisions. The hydrostatic Eqs.
(\ref{energy1}) should be supplemented by constitutive relations:
$p, \kappa$ and $I$ in terms of $n$ and $T$. These relations are
derivable systematically only in the dilute limit
\cite{hydro,dilute}. Being interested in moderate densities, we
shall employ the well-known constitutive relations by Jenkins and
Richman \cite{JR}, that account for excluded particle volume. In
the nearly-elastic limit one can neglect the inelasticity
correction terms in $p$ and $\kappa$, as well as the small density
gradient term, proportional to $1-r$, in the heat flux
\cite{Brey1}.

Equations (\ref{energy1}) can be rewritten in terms of a single
variable: the scaled inverse density $z(x,y)=n_c/n(x,y)$, where
$n_c= 2/(\sqrt{3} d^2)$ is the hexagonal close-packing density. In
scaled coordinates, ${\mathbf r} /L_x \to {\mathbf r}$, the box
dimensions become $1 \times \Delta$, where $\Delta = L_y/L_x$ is
the box aspect ratio. We obtain \cite{KM}
\begin{equation}
\nabla\cdot(F(z)\,\nabla z)=\eta\, Q(z)\,,
\end{equation}
where $F(z)= A(z)\,B(z)$,
\begin{eqnarray}
A(z)&=&\frac{G\left[1+\frac{9\pi}{16}\left(1+\frac{2}{3G}\right)^2\right]}{z^{1/2}(1+2G)^{5/2}}\,,\nonumber
\\
B(z)&=&1+2G+\frac{\pi}{\sqrt{3}}\frac{z(z+\frac{\pi}{16\sqrt{3}})}{(z-\frac{\pi}{2\sqrt{3}})^3}\,,\nonumber
\\
Q(z)&=&\frac{6}{\pi}\frac{z^{1/2}G}{(1+2G)^{3/2}}\,,\nonumber
\\
G&=&G(z)=\frac{\pi}{2\sqrt{3}}\frac{z-\frac{7\pi}{32\sqrt{3}}}{(z-\frac{\pi}{2\sqrt{3}})^2}\,,
\label{E5}
\end{eqnarray}
and $\eta=(2\pi/3)(1-r)(L_x/d)^2$ is the hydrodynamic inelasticity
parameter. Introducing
$\psi(x,y)=\int_0^{z}F(z^{\prime})\,dz^{\prime}$, we arrive at the
following equation:
\begin{equation}
\nabla^2\psi=\eta\,\tilde{Q}(\psi)\,, \label{F20}
\end{equation}
where $\tilde{Q}(\psi)=Q\left[z(\psi)\right]$ (in the following
the symbol ~$\tilde{}$~ is omitted). The boundary conditions  are
\begin{equation}
\left.\frac{\partial\psi}{\partial x}\right|_{x=0}=
\left.\frac{\partial\psi}{\partial y}\right|_{x=1}=
\left.\frac{\partial\psi}{\partial y}\right|_{y=-\Delta/2}=
\left.\frac{\partial\psi}{\partial y}\right|_{y=\Delta/2}=0\,.
\label{F30}
\end{equation}
Finally, the number of particles is conserved:
\begin{equation}
\frac{1}{\Delta}\,
\int\limits_{\hspace{-0.5mm}-\Delta/2}^{\Delta/2}\!\,\int\limits_0^1\frac{dx
dy }{z(\psi)}= \frac{N}{L_x L_y n_c} \equiv f \, . \label{F60}
\end{equation}
The hydrostatic problem (\ref{F20})-(\ref{F60}) is fully
determined by three scaled parameters: the area fraction $f$,
$\eta$, and $\Delta$. Notice that the steady-state
\textit{density} distributions are independent of $T_0$, as the
hard sphere model does not introduce any intrinsic energy scale.

\section{Stripe state, symmetry-breaking
instability and bifurcation curve} The trivial steady state of the
system is a laterally uniform cluster of particles located at the
wall $x=0$, opposite to the thermal wall \cite{Grossman}, see
Fig.~1. This state will be called the stripe state.  In the
language of hydrodynamics, it is described by the $y$-independent
solution of Eqs. (\ref{F20})-(\ref{F60}); we shall denote it by
$z=Z(x)$, correspondingly $\psi=\Psi(x)$.
\begin{figure}[ht]
\epsfig{file=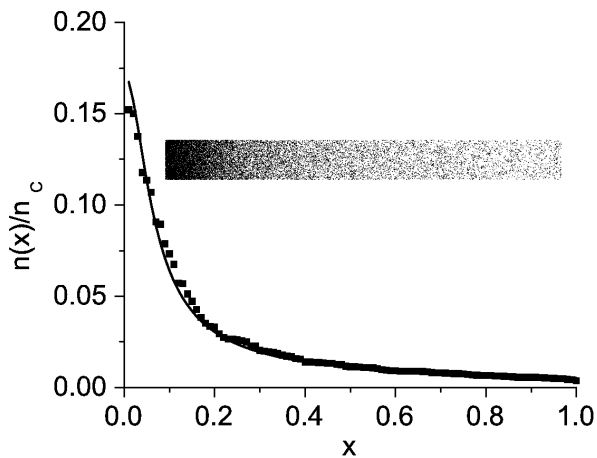, width=7.8cm, clip=} \caption{The stripe
state for $\eta=11,050$ and $f=0.025$. We show the scaled density
versus scaled coordinate $x$ obtained (a) by solving numerically
Eqs. (\ref{F20}) -(\ref{F60}) in one dimension (line) and (b) in
EMD simulation with $N=2 \cdot 10^4$ particles for $\Delta=0.1$
(squares). The inset shows a snapshot of the system from the EMD
simulation (the hot wall is on the right). Because of a finite
image resolution the particle number density in this and other
snapshots may look higher than it is.} \label{fig1}
\end{figure}

It was predicted that, in a wide region of the parameter space
$(f,\eta,\Delta)$, the stripe state should give way, by a
symmetry-breaking bifurcation (either supercritical or
subcritical), to a laterally asymmetric state
\cite{LMS,Brey2,KM,LMS2}. For very large aspect ratios $\Delta$,
this phase-separation instability has been observed in EMD
simulations \cite{Argentina}. For a laterally asymmetric steady
state one can write
\begin{equation}
\psi(x,y)=\Psi(x) + \sum\limits_n\, \varphi_n(x)\, \exp (inky)\,,
\label{F180}
\end{equation}
where $\varphi_{-n}(x)=\varphi^*_{n}(x)$. What happens close to
the supercritical bifurcation point? Here the leading terms are
those with $n = \pm 1$, while $\varphi_0\sim\varphi_1^2$,
$\varphi_2\sim\varphi_1^2$, $\varphi_3\sim\varphi_1^3$, etc. The
bifurcation point itself can be found from the linear eigenvalue
problem
\begin{equation}
\varphi_{1k}^{\prime\prime}-\eta\, Q_{\Psi}\, \varphi_{1k} -
k_c^2\,\varphi_{1k}=0\, , \label{F130}
\end{equation}
\begin{equation}
\varphi_{1k}^{\prime}(0)=0\,\,\, \mbox{and}\,\,\, \varphi_{1k}
(1)=0 \label{F140}
\end{equation}
that was analyzed in Refs. \cite{LMS,Brey2,KM}. Here
$$Q_\Psi(x)=
\left.F^{-1}\,dQ/dz\right|_{z=Z(x)}.$$ For given $\eta$ and $f$,
one obtains the eigenvalue $k=k_c(\eta,f)$ and corresponding
eigenfunction $\varphi_{1k}(x)$. The modes with $k<k_c(\eta,f)$
are unstable. Within a spinodal interval $f_1(\eta)<f<f_2(\eta)$,
the effective lateral compressibility of the gas is
\textit{negative}, and this is the mechanism of the instability
\cite{KM,Argentina}. At $\eta \gg 1$, there is a range of $f$ such
that $k_c$ and $\varphi_{1k}(x)$ become insensitive to the precise
form of the boundary conditions at the driving wall. This is the
universal "localization regime", when the eigenfunction
$\varphi_{1k}(x)$ is exponentially localized at the wall opposite
to the driving wall \cite{LMS,KM}. The spinodal interval exists
for $\eta_c<\eta<\infty$; it shrinks to zero at $\eta=\eta_c
\simeq 344.3$ \cite{Argentina,KMS}. It has been recently shown,
for a different boundary condition at the driving wall, that the
bifurcation from the stripe state to a phase-separated state is
supercritical within some density interval
$f_{-}(\eta)<f<f_{+}(\eta)$, which is located within the spinodal
interval.  On each of the intervals $f_1<f<f_{-}$ and
$f_{+}<f<f_2$,  the bifurcation is subcritical~\cite{LMS2}.

As we have already noted, the present work focuses on the phase
separation via a supercritical bifurcation. To obtain the
asymptotics of the supercritical bifurcation curve close to onset,
one should go to the second order of the perturbation theory and
take into account, in Eq. (\ref{F180}), the terms $n=0,\pm 1$ and
$\pm 2$. In this way one obtains three linear ordinary
differential equations, presented in Ref. \cite{LMS2}, where the
same problem was solved for a different boundary condition at the
driving wall. The solvability condition for these equations
\cite{Iooss} yields the bifurcation curve: $A$ versus $k_c^2-k^2$.
The amplitude $A$ can be uniquely defined by the relation
$$\varphi
(x) = A \,\Phi_0(x)+ A|A|^2\, \delta\varphi (x)\,,$$ where
$\Phi_0(x)$ is the solution of Eqs. (\ref{F130}) and (\ref{F140})
such that $\Phi_0(0)=1$, while $\delta\varphi (x)=\mathcal{O}(1)$.
This yields
$$A\left(k_c^2-k^2\right)=CA|A|^2\,,$$
where $C=$const.
The trivial solution $A=0$ describes the stripe state, while the
nontrivial one, $k_c^2-k^2=C|A|^2 $ describes the bifurcated
state. The constant $C$ can be computed numerically. $C>0 \, (<0)$
corresponds to supercritical (subcritical) bifurcation. We present
here the resulting bifurcation curve for $Y_c$, the (normalized)
$y$-coordinate of the center of mass of the granulate
\begin{equation}
Y_c= \frac{\int_0^1 dx \int_{-\Delta/2}^{\Delta/2} \, y\, n(x,y)\,
dy }{\Delta \int_0^1 dx \int_{-\Delta/2}^{\Delta/2} n(x,y)\, dy
}\,, \label{cmass}
\end{equation}
Let us fix $\eta$ and $f$ and treat $\Delta$ as the control
parameter. When $\Delta$ is slightly larger than
$\Delta_c=\pi/k_c(f)$, only the fundamental mode $k=\pi/\Delta$ is
unstable, and the bifurcation curve has the form
\begin{equation}
\left|Y_c\right|= \Upsilon \left(\Delta-\Delta_c\right)^{1/2}\, .
\label{MT420}
\end{equation}
Here $$\Upsilon=\frac{2^{3/2} f_0}{C^{1/2} \Delta_c
f}\,,\,\,\,\,\,\, f_0=2\int_0^1 dx \,\frac{\Phi_{01}}{ Z^2 F}\,,$$
and $\Phi_{01}(x)$ is the solution of initial-value problem for
Eq. (\ref{F130}) with the initial conditions $Y(0)=1$ and
$Y^{\prime}(0)=0$. Equation (\ref{MT420}) assumes $C>0$: a
supercritical bifurcation. We have found that, at fixed $\eta$,
$C>0$ on an interval $f_{-}(\eta)<f<f_{+}(\eta)$ that lies within
the spinodal interval $(f_1,f_2)$. On the intervals $f_1<f<f_{-}$
and $f_{+}<f<f_2$ the coefficient $C$ becomes negative which
indicates a subcritical bifurcation. The solid line in Fig.~6
shows the supercritical bifurcation curve (\ref{MT420}) for
$\eta=11,050$ and $f=0.025$. Here $\Delta_c \simeq 0.514$ and
$\Upsilon \simeq 0.142$.

When $\Delta$ is well above $\Delta_c$, the weakly nonlinear
theory is invalid, and a numerical solution of the fully nonlinear
hydrostatic problem  (\ref{F20})-(\ref{F60}) is needed for the
determination of $\left|Y_c\right|$. An alternative approach is a
hydrodynamic simulation, that is a numerical solution of the
hydrodynamic equations. Numerical simulations of this type were
done in Ref. \cite{LMS2} for a different version of constitutive
relations \cite{Grossman} and a different boundary condition at
the driving wall. It was observed that the phase-separation
instability produces multiple clusters whose further dynamics
proceed as gas-mediated competition and coarsening. Direct merging
of clusters can also occur. The final symmetry-broken state, as
observed in the hydrodynamic simulations, is always a single,
almost densely packed stationary 2D cluster coexisting with gas
(or dilute bubble coexisting with denser fluid). The cluster is
located in one of the system's corners (unless periodic boundary
conditions are used). This scenario was confirmed in a
hydrodynamic simulation of the \textit{present} system (for
$\eta=11,050$, $f=0.025$ and $\Delta=3$) done by E. Livne
\cite{Livne}. A density map of the hydrodynamic final state in
this case is shown in Fig.~2D. The steady-state value
$\left|Y_c\right| \simeq 0.265$, obtained in this simulation, is
shown by the circle in Fig.~6.

\section{EMD simulations}

\subsection{Simulation method, parameters and diagnostics}

We put the predictions of the granular hydrostatics into test by
doing extensive EMD simulations of this system. Most of the
simulations were done with $N=2\cdot 10^4$ particles: hard disks
of diameter $d=1$ and mass $m=1$. The thermal wall temperature is
$T_0=1$, so the scaled time unit is $d\,(m/T_0)^{1/2}=1$. A
standard event-driven algorithm \cite{Alder} was used. Two of the
hydrodynamic parameters, $\eta=11,050$ and $f=0.025$, were fixed
in all simulations, while $\Delta$ was varied in the range of
$0.1<\Delta<3$. This was achieved by varying $L_x$, $L_y$ and $r$.
Indeed, for a fixed $\eta$, $f$, $\Delta$ and $N$ the coefficient
of normal restitution,
\begin{equation} \label{restitution}
r = 1-\frac{\sqrt{3}\, \eta f \Delta}{\pi N}\,,
\end{equation}
and the system's dimensions,
\begin{equation} \label{dimensions}
L_x = \left(\frac{\sqrt{3}\, N}{2 f \Delta}\right)^{1/2}\,\,\,
\mbox{and} \,\,\,L_y=\left(\frac{\sqrt{3}\, N \Delta}{2
f}\right)^{1/2}\,,
\end{equation}
are uniquely determined. For the values of the parameters that we
used, $r$ was always in the range of nearly elastic collisions: $r
\geq 0.977$. The initial spatial distribution of the particles was
(statistically) uniform, while the initial velocity distribution
was Maxwell's with the wall temperature $T_0=1$. The
center-of-mass coordinate $Y_c (t)$ was used as a quantitative
probe of the lateral asymmetry of the system. Before taking the
steady-state measurements we waited until transients died out.
This was monitored by the time-dependence of the average kinetic
energy of the particles (that first decayed and then approached an
almost constant value) and by the time-dependence of the center of
mass itself, see below. Selected movies of these simulations can
be downloaded from
http://bioinf.charite.de/kies/giantfluctuations/.

\subsection{Final states at different $\Delta$}

The EMD simulations showed that, at aspect ratios \textit{well
below} the threshold value of $\Delta=\Delta_c\simeq 0.512$, the
final state is a (weakly fluctuating) stripe state. The number
density profile versus $x$, found in the simulations, compares
very well with the hydrostatic solution (see Fig.~1), while $Y_c
(t)$ stays close to zero. Notice that the Jenkins-Richman
constitutive relations \cite{JR}, that we used in this comparison,
do not include any fitting parameters. Therefore, \textit{well
below} the instability threshold in $\Delta$, the hydrostatic
solution yields a quantitatively accurate leading-order
description of the system.

At aspect ratios \textit{well above} the instability threshold we
always observed several clusters nucleating at the wall opposite
to the driving wall. The cluster dynamics (Fig.~2 A to C) proceeds
as gas-mediated competition and coarsening (sometimes as direct
mergers) of clusters, in accord with hydrodynamic simulations
\cite{LMS2}. As time increases, the number of clusters goes down,
and only one dense cluster, fluctuating around its average
position in one of the two corners, opposite to the thermal wall,
finally survives. Fig.~2C shows a snapshot of the final state for
$\Delta=3$. For comparison, Fig.~2D shows a density map of the
final steady state obtained by E. Livne in a \textit{hydrodynamic}
simulation for the same hydrodynamic parameters. The
center-of-mass position $Y_c$ of the steady state agrees well with
the average-in-time center-of-mass position, measured in the EMD
simulations, as shown by the circle in Fig.~6. This indicates
that, \textit{well above} the instability threshold, the
hydrostatic theory describes the steady states of the system well.
We can also refer the reader to the recent EMD simulation results
for \textit{very} large aspect ratios \cite{Argentina}. As no
appreciable fluctuations around a broken-symmetry steady state
were reported, one can safely assume that the broken-symmetry
steady states observed in Ref. \cite{Argentina} should be also
describable by the hydrostatic theory.

\begin{figure}[hb!]
\begin{tabular}{cccc}
 A &\hspace{1mm} B &\hspace{1mm} C &\hspace{1mm} D\\[0.3cm]
 \epsfxsize=1.9cm  \epsffile{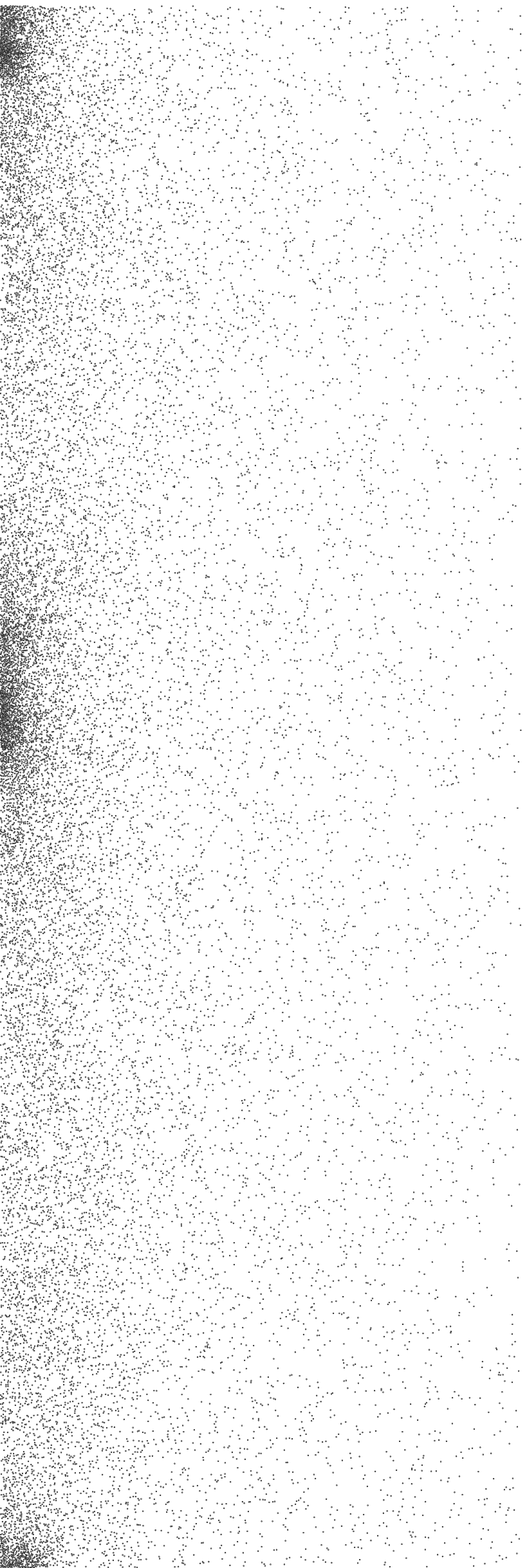} &
 \hspace{1mm} \epsfxsize=1.9cm  \epsffile{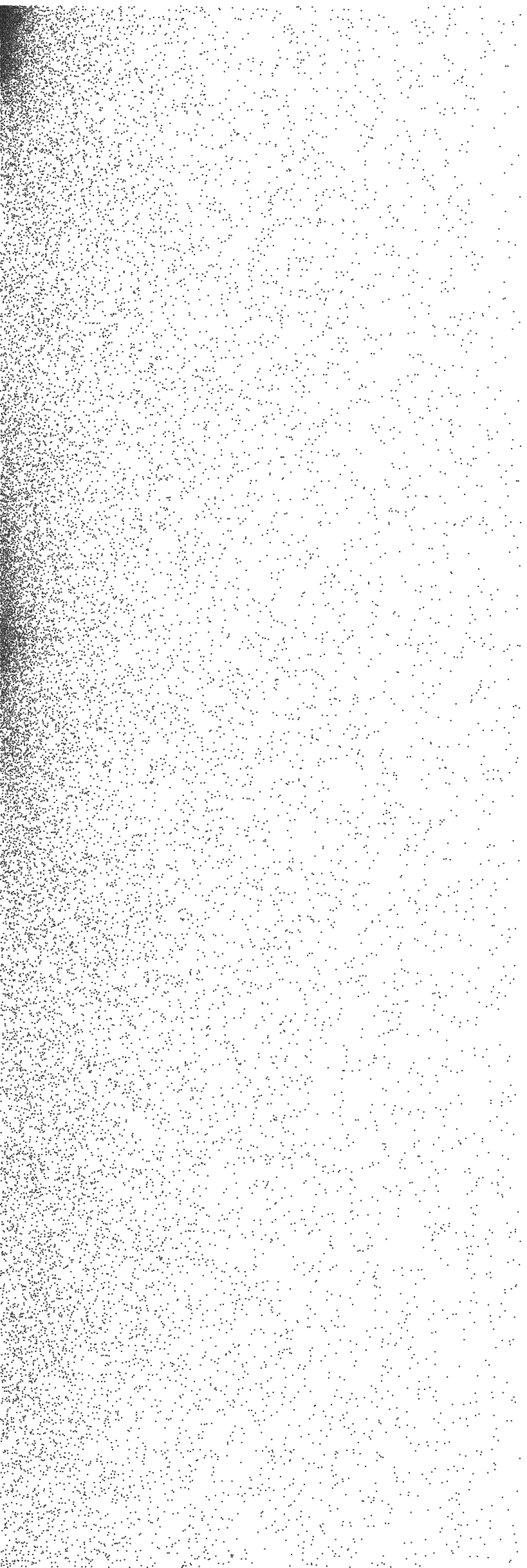} &
 \hspace{1mm} \epsfxsize=1.9cm  \epsffile{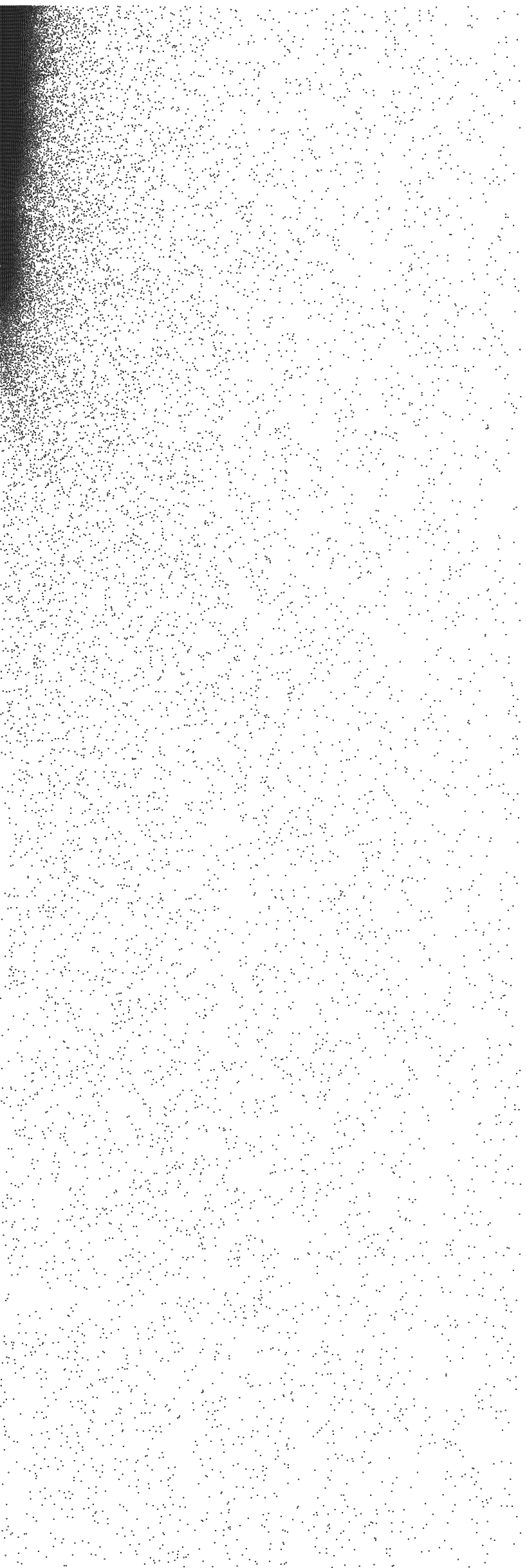} &
 \hspace{1mm} \epsfxsize=1.9cm  \vspace{10mm}\epsffile{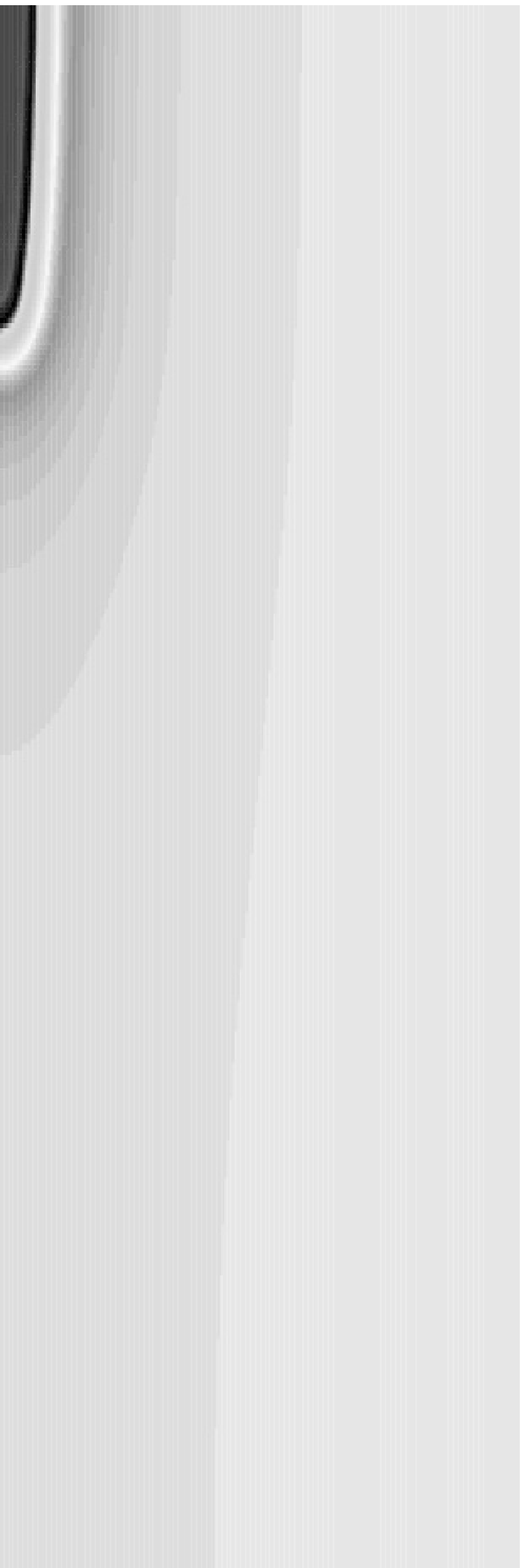}\\
\end{tabular}
\caption{Nucleation and coarsening of clusters as observed in an
EMD simulation with $N = 2 \cdot 10^4$ particles for
$\eta=11,050$, $f=0.025$ and $\Delta=3$. The hot wall is on the
right.  The scaled times are $14,425$ (A), $26,218$ (B) and
$191,616$ (C). Figure D is a density map of the steady state
obtained by E. Livne in a simplified \textit{hydrodynamic}
simulation for the same hydrodynamic parameters \cite{Livne}. }
\label{fig2}
\end{figure}

The system behavior changes dramatically, however, as the aspect
ratio $\Delta$ approaches $\Delta_c$. We found that, in a
\textit{wide} region of $\Delta$ around $\Delta_c$, the final
state of the system exhibits large-amplitude irregular
oscillations, as dense clusters at the wall opposite to the
driving wall nucleate, move in the lateral direction, dissolve and
reappear. Figure 3 shows a typical sequence of snapshots from an
EMD simulation for $\Delta=1$.

\begin{figure}[hb!]
\centerline{
  \includegraphics[width=2.7cm,angle=180]{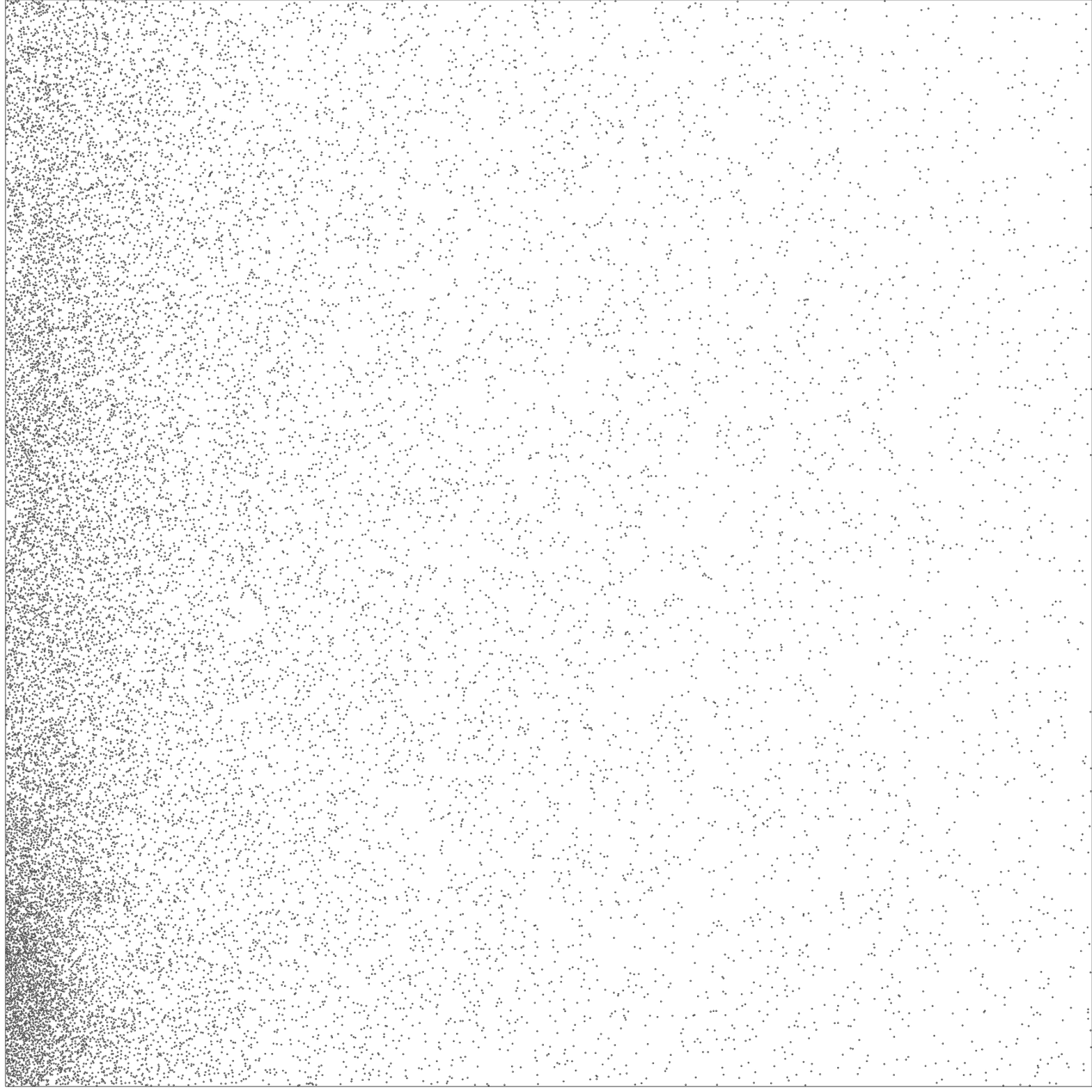}
  \includegraphics[width=2.7cm,angle=180]{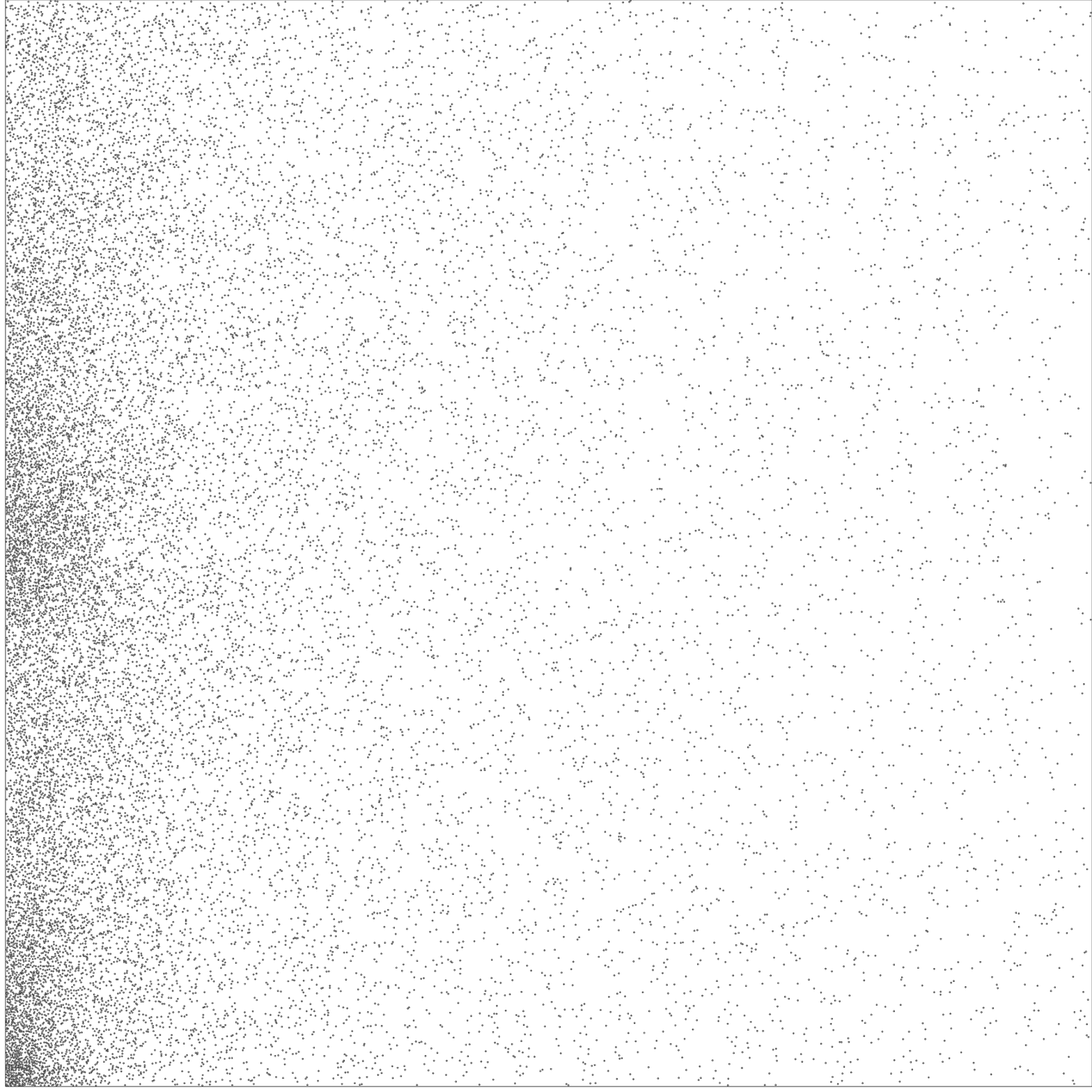}
  \includegraphics[width=2.7cm,angle=180]{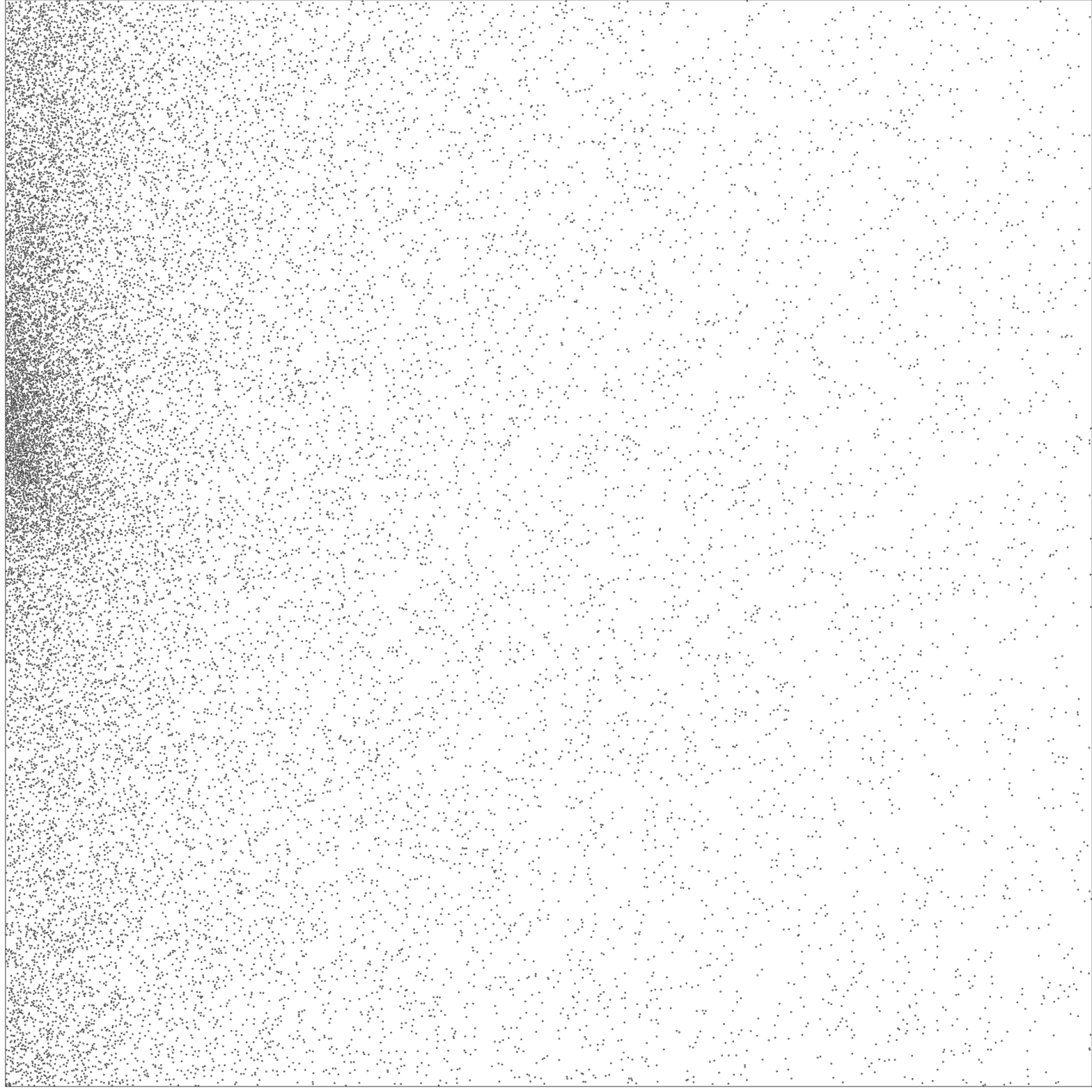}
} \vspace*{0.1cm} \centerline{
  \includegraphics[width=2.7cm,angle=180]{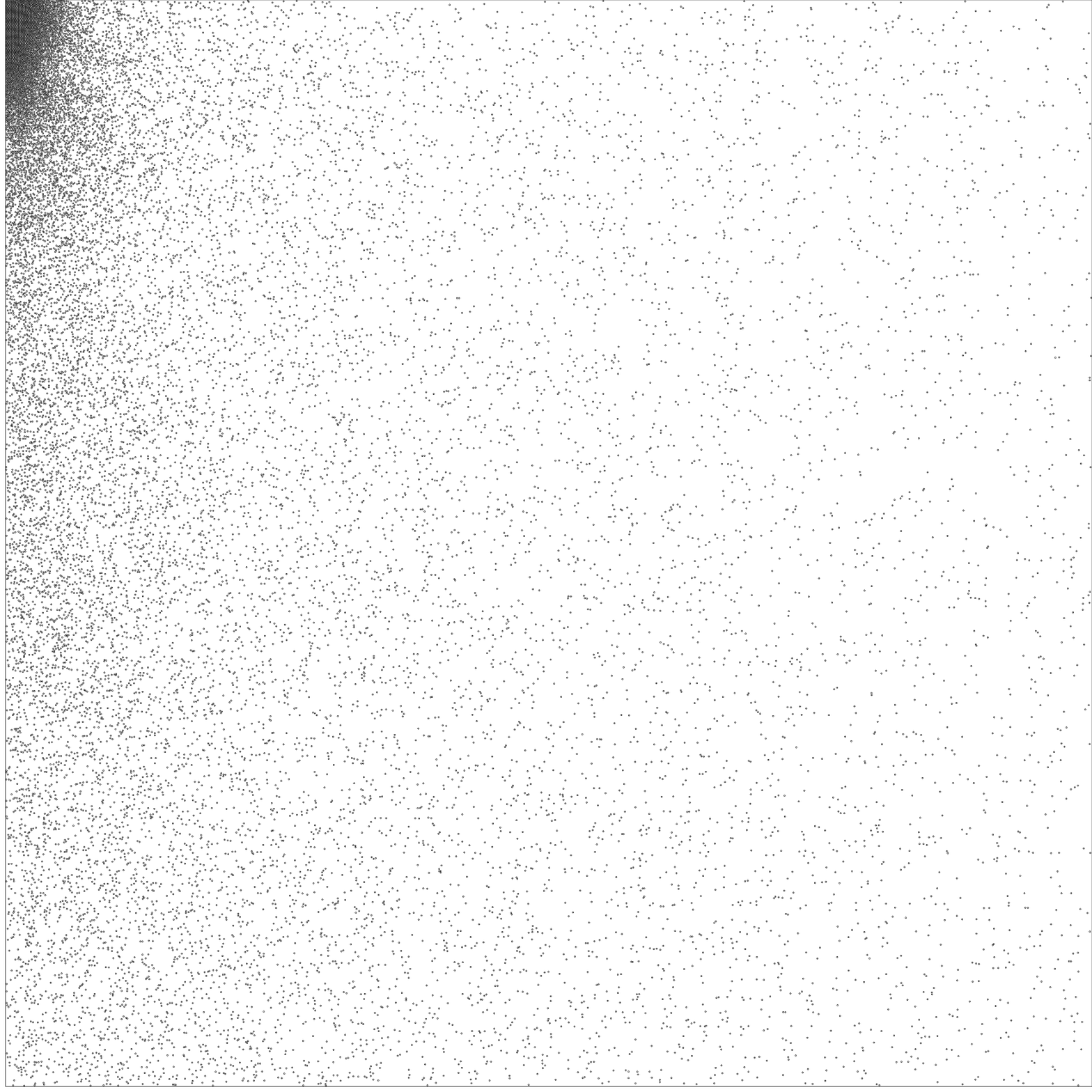}
  \includegraphics[width=2.7cm,angle=180]{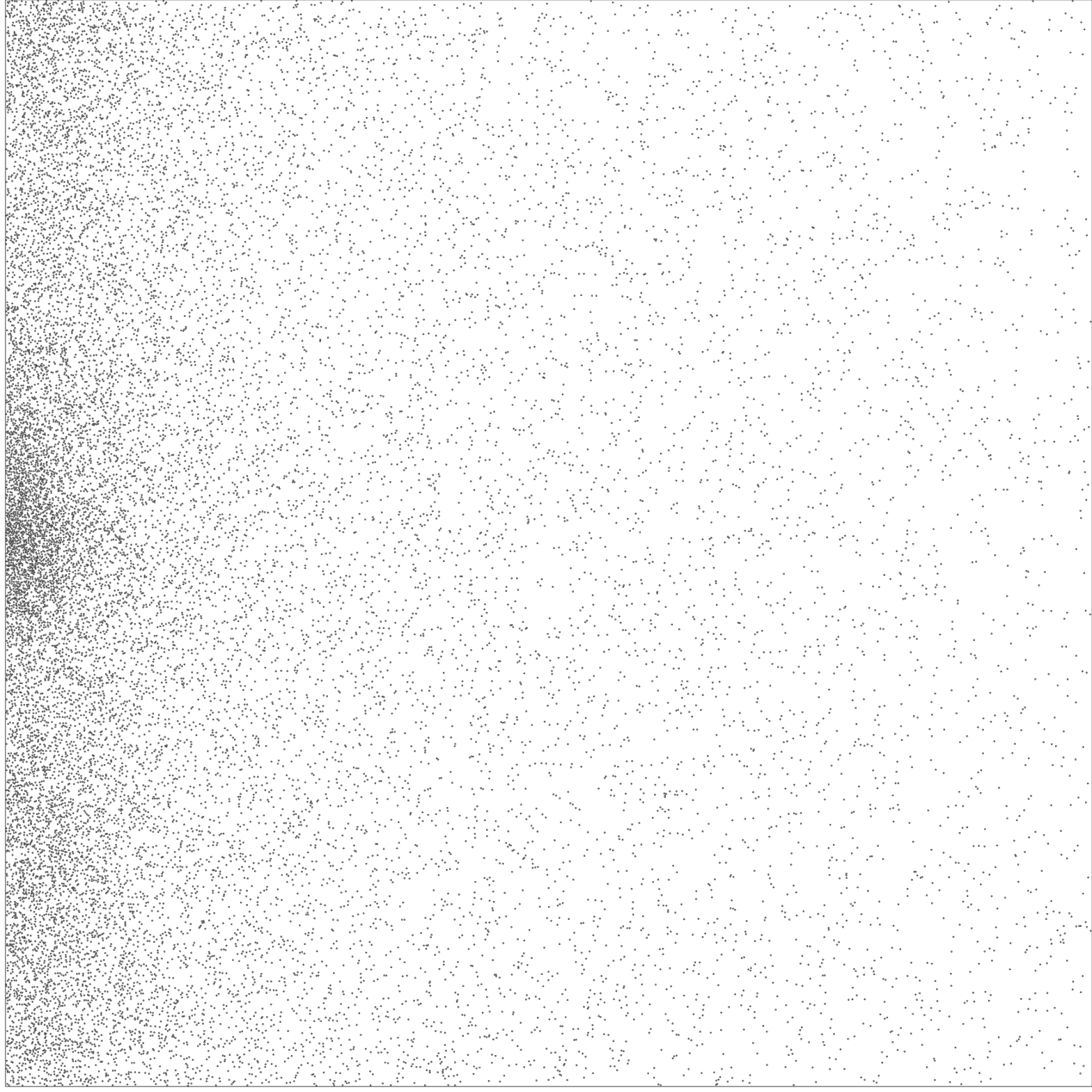}
  \includegraphics[width=2.7cm,angle=180]{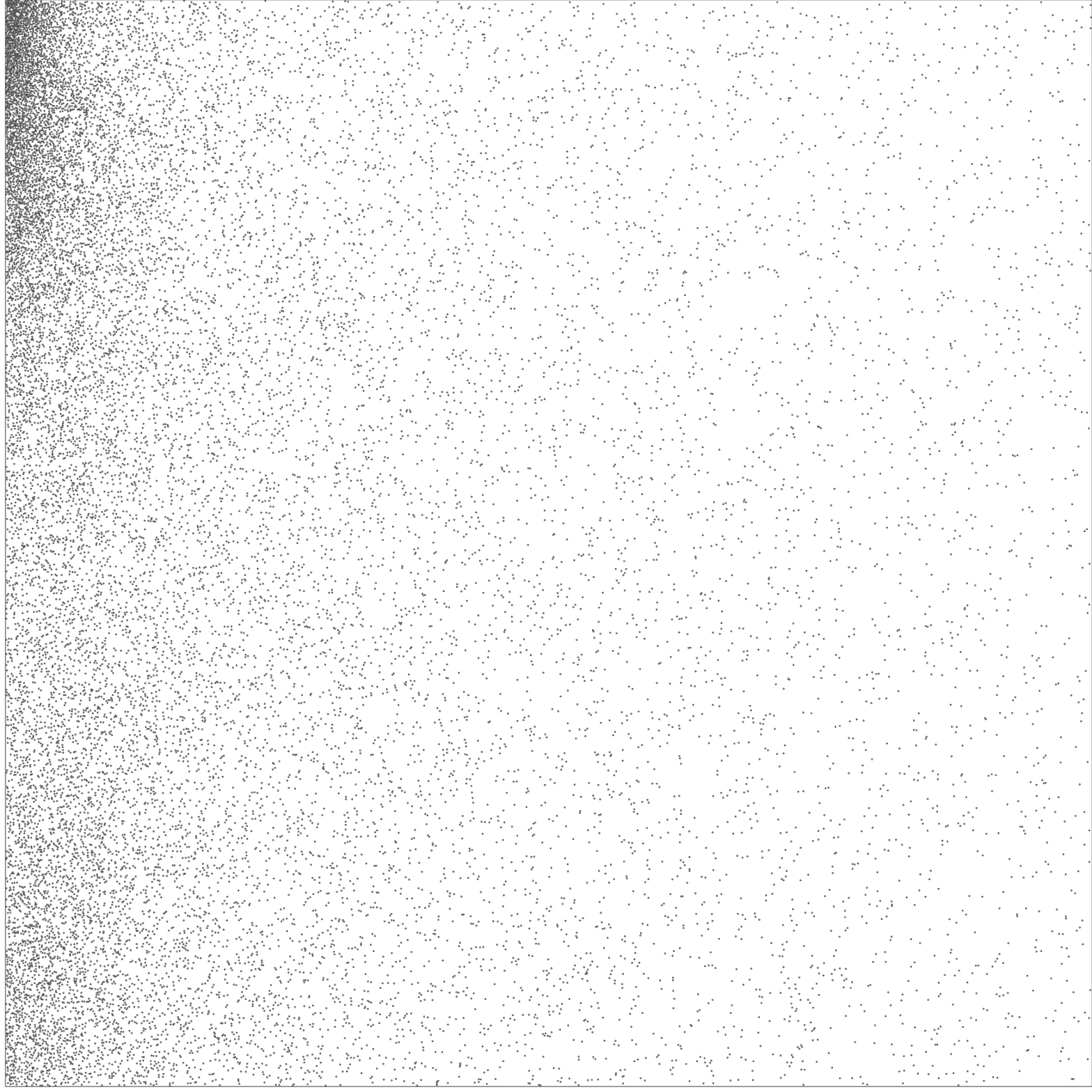}
}
\caption{Irregular lateral cluster dynamics for $\Delta=1$ as
observed in an EMD simulation with $N = 2 \cdot 10^4$ particles
for $\eta=11,050$ and $f=0.025$. The time progresses from left to
right, starting from the upper row. The hot wall is on the right.}
\label{fig3}
\end{figure}

\begin{figure}[hb!]
\begin{tabular}{cc}
  \epsfxsize=4.0cm  \epsffile{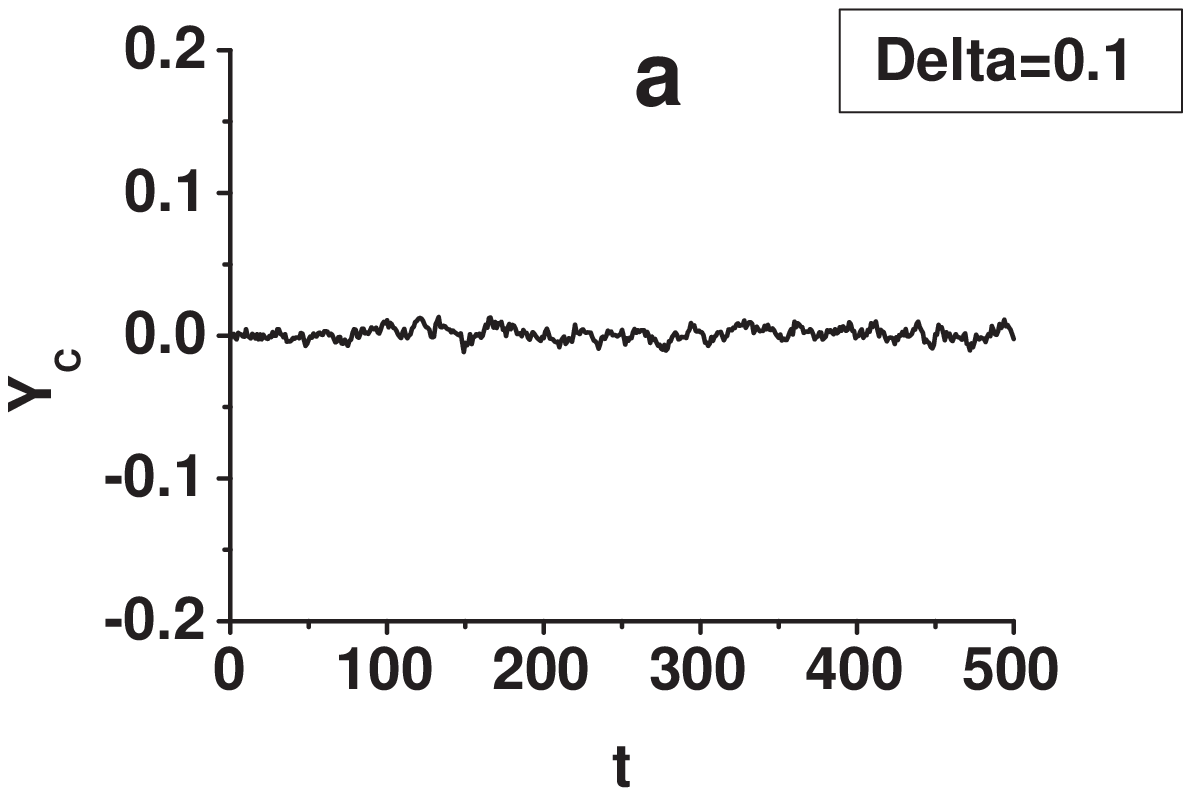} & \epsfxsize=4.0cm  \epsffile{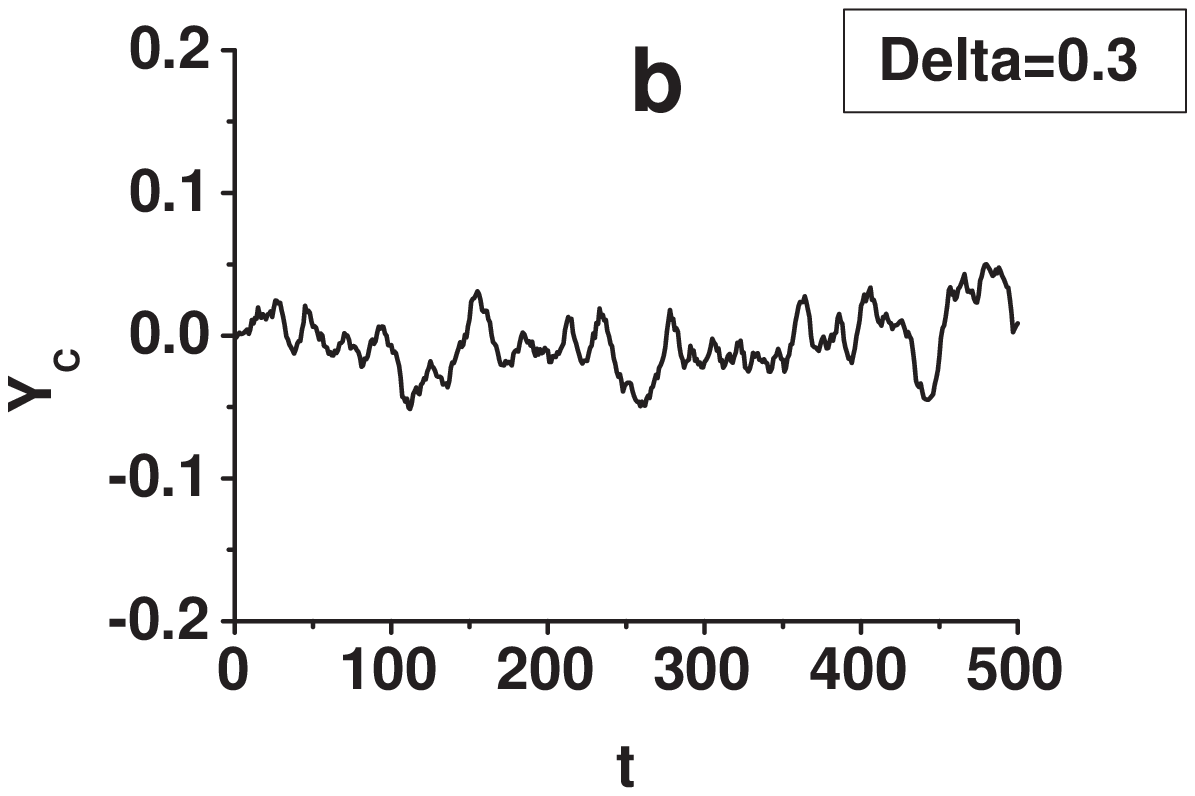}\\
  \epsfxsize=4.0cm  \epsffile{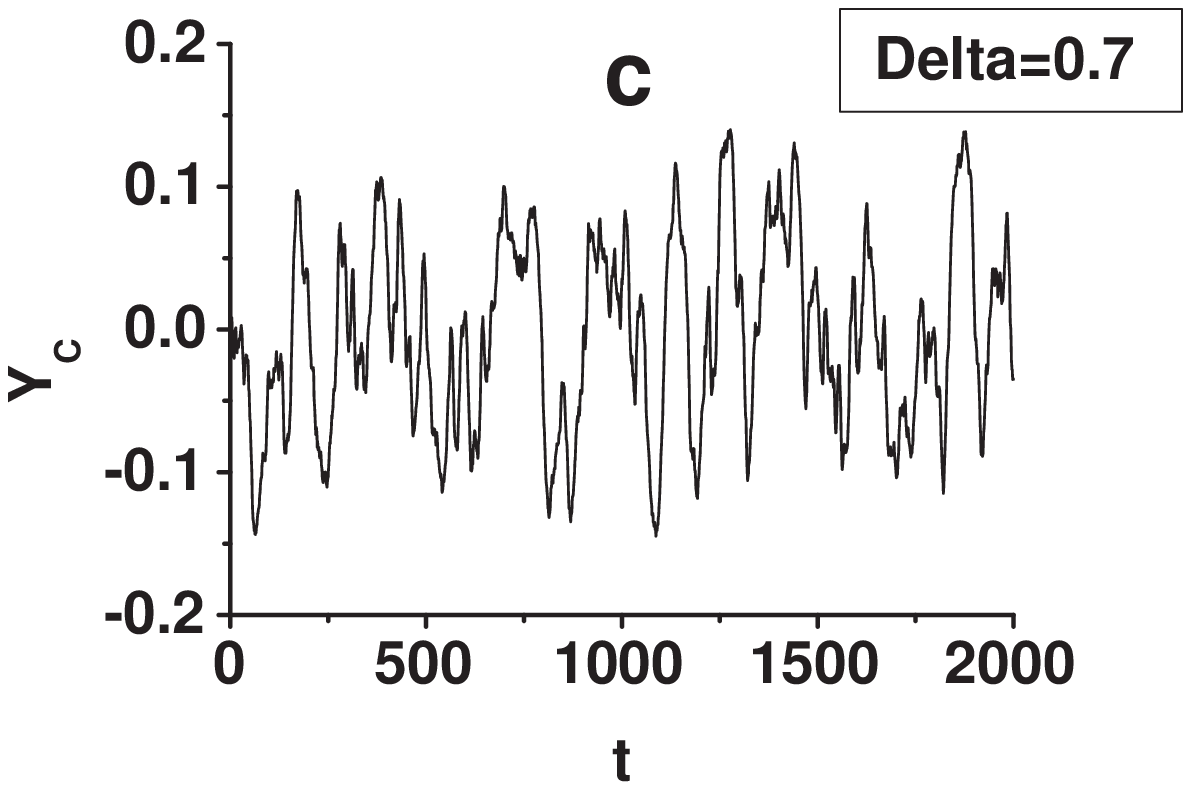} & \epsfxsize=4.0cm  \epsffile{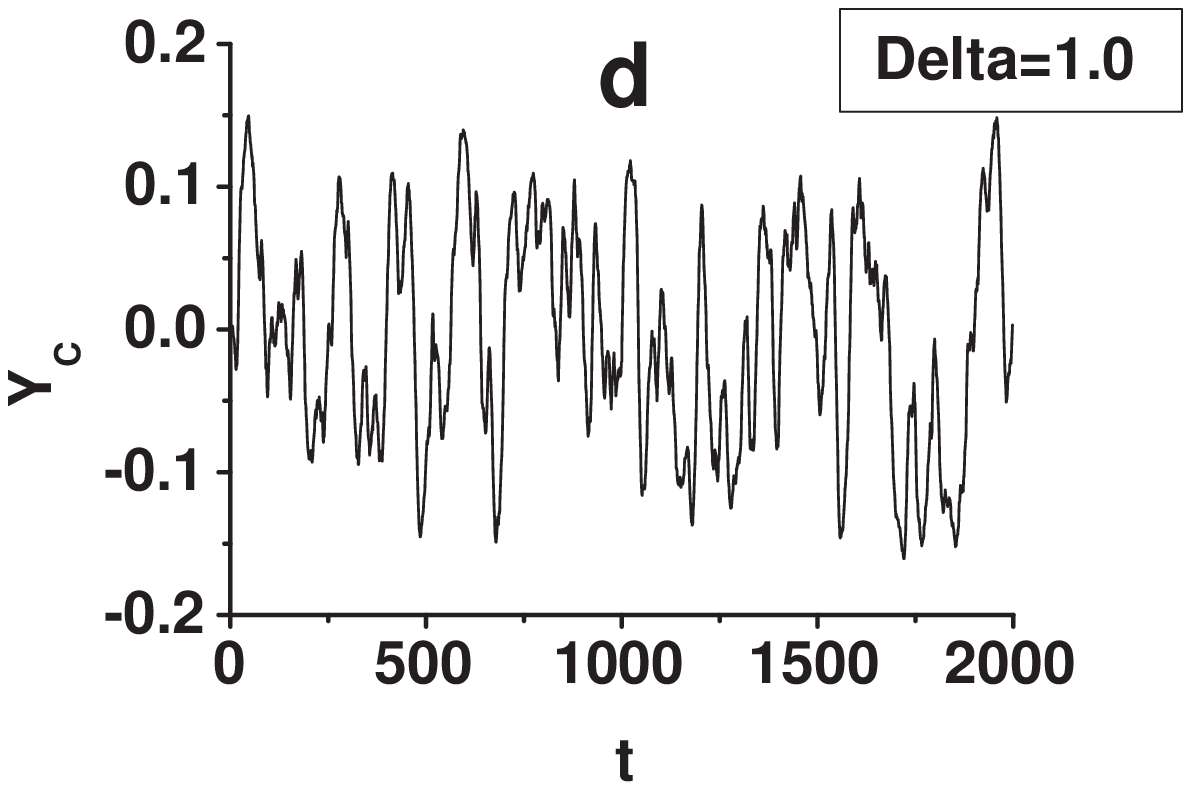}\\
  \epsfxsize=4.0cm  \epsffile{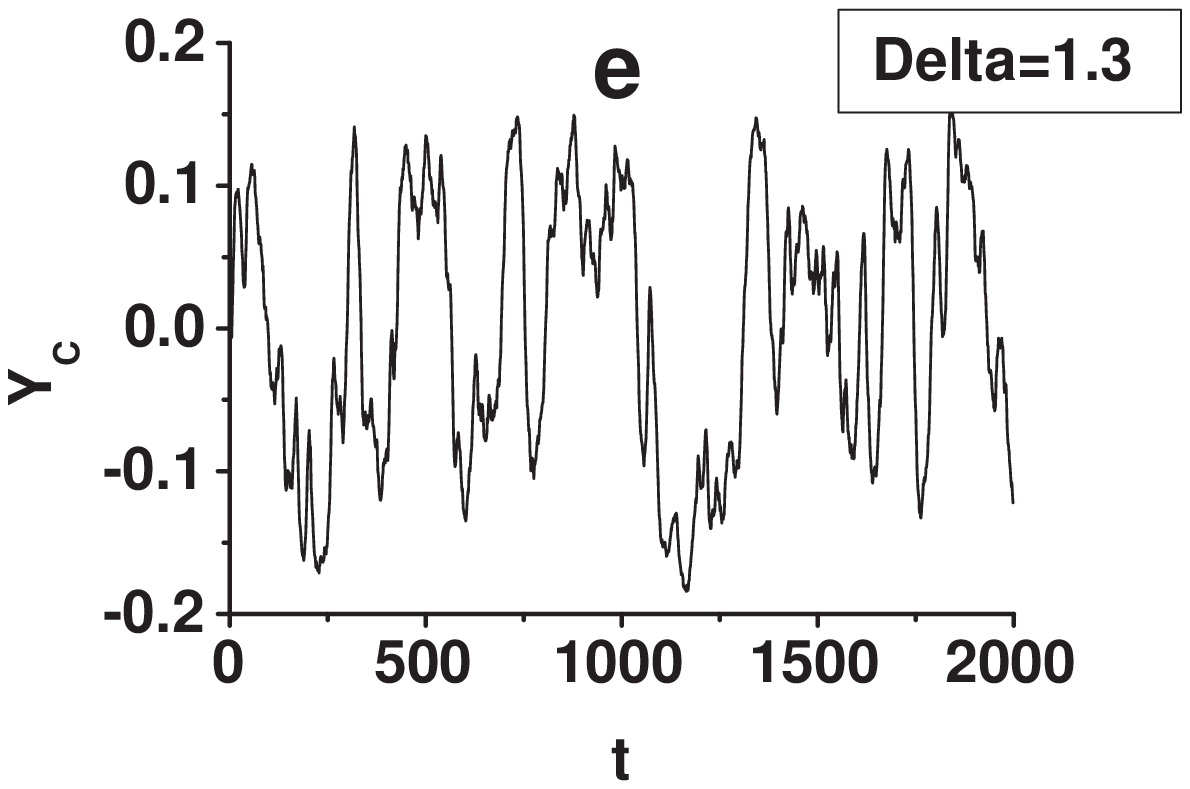} & \epsfxsize=4.0cm  \epsffile{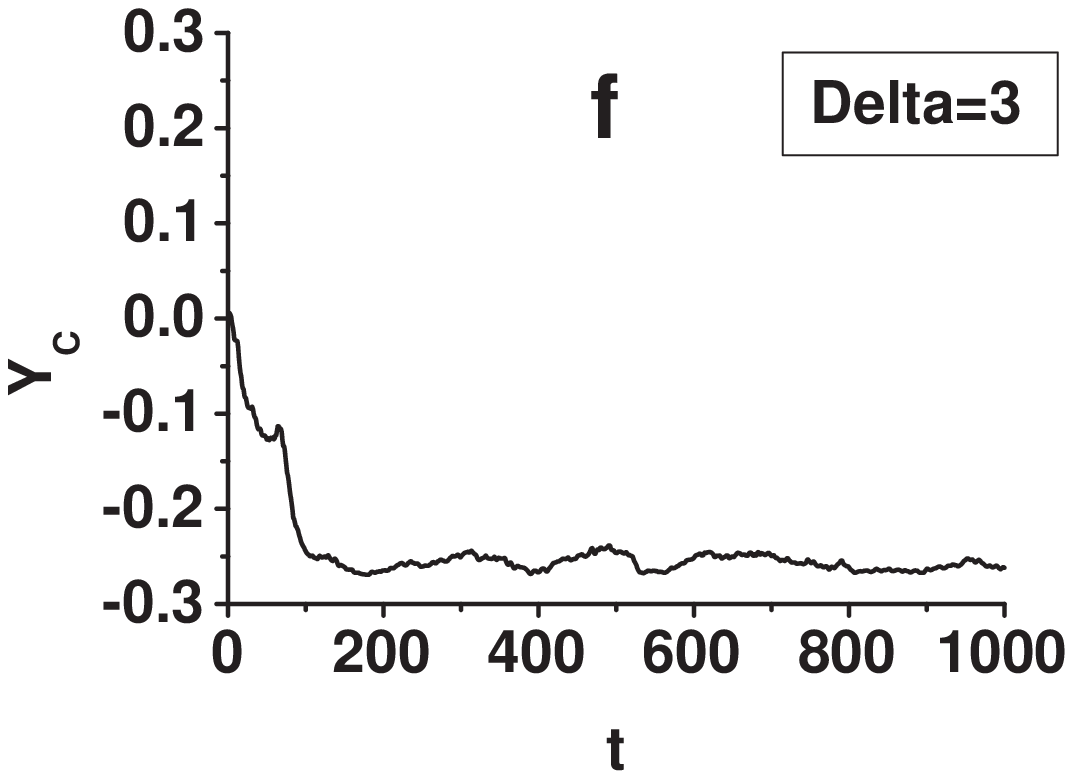}
\end{tabular}
\caption{$Y_c$ versus time for $\eta=11,050$ and $f=0.025$ and
different values of the aspect ratio $\Delta$, as observed in EMD
simulations with $N=2 \cdot 10^4$ particles. Time here is
proportional to the number of particle collisions; $t=500$
corresponds to $505,036$ scaled time units.} \label{fig4}
\end{figure}

Figure 4 shows the time history of the center-of-mass coordinate
$Y_c$ for six different values of $\Delta$.  One can see that, in
a wide region of intermediate $\Delta$, the center-of-mass
coordinate $Y_c(t)$ shows large-amplitude irregular oscillations.
Noticeable are multiple zero crossings of $Y_c(t)$ at aspect
ratios \textit{above} the hydrodynamic bifurcation point $\Delta_c
\simeq 0.512$ (Fig.~4 c-e). Smaller but still significant
irregular oscillations are also observed \textit{below}
$\Delta_c$, as if the system persistently tends to break the
lateral symmetry there. The hydrostatic picture is recovered when
one moves farther away, in any direction, from the region of
$\Delta \sim \Delta_c$. Indeed, Fig. 4e shows that zero crossings
of $Y_c (t)$ occur less often for $\Delta=1.3$, than for $\Delta =
0.7$ or $1$. At still larger $\Delta$ (Fig.~4f) no zero crossings
are observed for any reasonable simulation time, and $Y_c$
fluctuates around a constant value that is very close to that
predicted by the hydrostatic theory (and shown by the circle in
Fig.~6).

To better characterize the fluctuation-dominated region, we
computed the probability distribution function $P(|Y_c|)$ of
different values of $|Y_c|$ in a statistical steady state, that
is, after transients die out. The stationarity of the remaining
data was tested by dividing the respective time interval into
three sub-intervals and  checking that the differences in
$P(|Y_c|)$ for the sub-intervals are small and not systematic. The
probability distribution $P(|Y_c|)$ is shown, at different
$\Delta$, in Fig.~5. At $\Delta \ll \Delta_c$ the maximum of
$P(|Y_c|)$ is at $|Y_c|=0$, and it is relatively narrow.
Correspondingly, there is no symmetry breaking there, the
fluctuations are relatively small, and the hydrostatic theory
yields an accurate leading-order description. At $\Delta \gg
\Delta_c$, the maximum of $P(|Y_c|)$ is at a non-zero $|Y_c|$.
This is a clear manifestation of symmetry-breaking: a dense
cluster develops in one of the corners away from the driving wall.
The probability distribution $P(|Y_c|)$ is also quite narrow here,
the fluctuations are relatively small, and there is a good
agreement between the hydrostatic theory and EMD-simulations. On
the contrary, in a wide region of $\Delta$ around $\Delta_c$, the
probability distribution $P(|Y_c|)$ is very broad, and the
hydrostatic theory breaks down. By following the position of the
maximum of $P(|Y_c|)$ at different $\Delta$ (see Fig.~6), one can
see that the symmetry-breaking transition occurs somewhere in the
region of $0.3 < \Delta < 1.0$. Because of the extreme flatness
and broadness of the probability distribution $P(|Y_c|)$ in this
region, a more accurate estimate of the position of the maximum of
$P(|Y_c|)$ requires a much better statistics (that is, a much
longer simulation time) than we could afford in this series of
simulations \cite{simulationtime}.

\begin{figure}[ht]
\epsfig{file=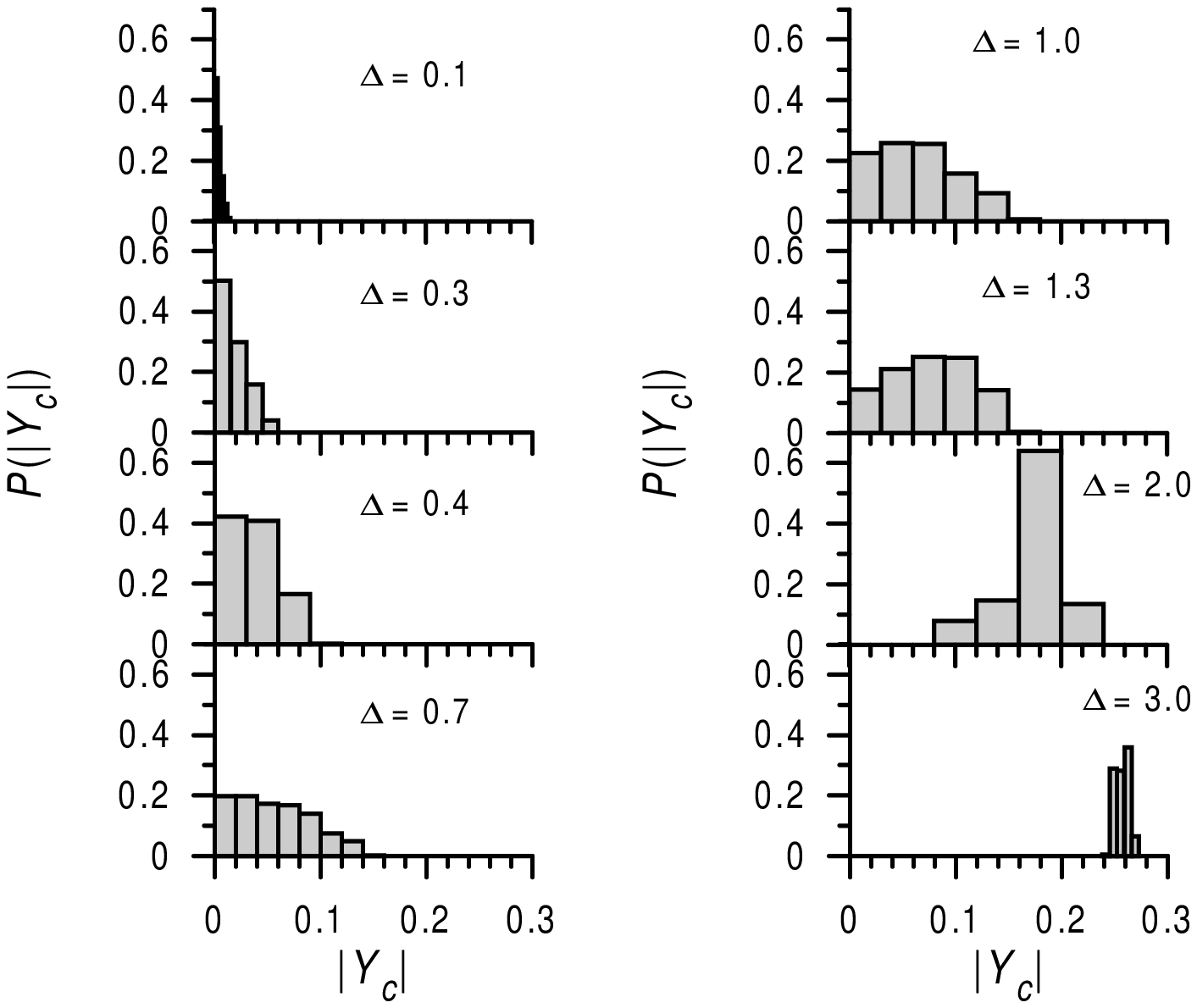, width=8.6cm, clip=} \caption{The
probability distribution function $P(|Y_c|)$ of the fluctuating
final state of the system for $\eta=11,050$ and $f=0.025$ and
different values of the aspect ratio $\Delta$, as observed in EMD
simulations with $N=2 \cdot 10^4$ particles. In order to show all
the graphs on the same scale, the probabilities (rather than the
probability densities) for each bin are shown.} \label{fig5}
\end{figure}

\begin{figure}[ht]
\epsfig{file=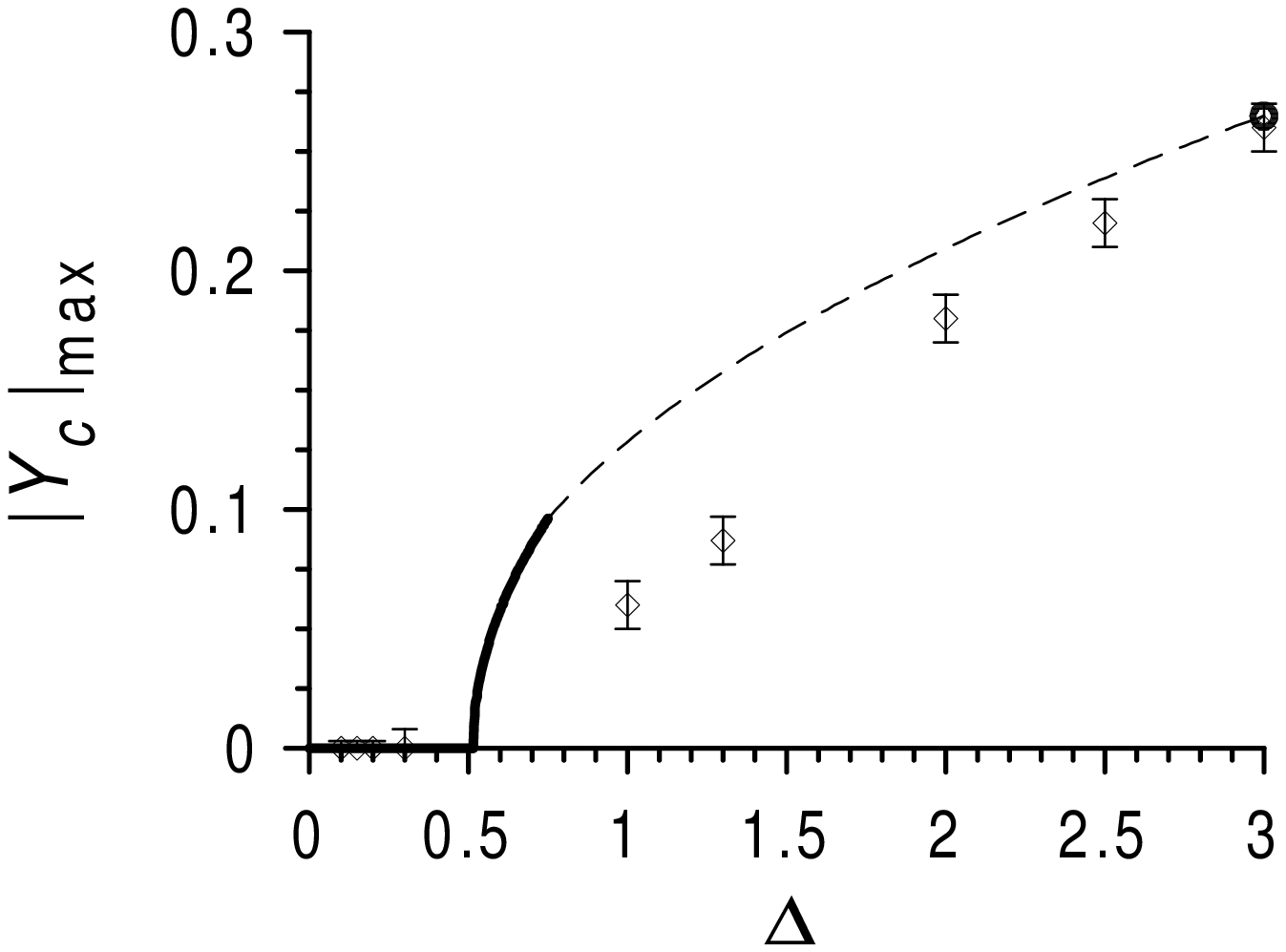, width=8.3cm, clip=} \caption{The
effective bifurcation diagram of the system for $\eta=11,050$ and
$f=0.025$, observed in EMD simulations with with $N=2 \cdot 10^4$
particles. Diamonds show, for each $\Delta$, the positions of the
maxima of the probability distribution function $P(|Y_c|)$. Above
the transition, the error bars show the errors in the estimation
of the position of the maximum of $P(|Y_c|)$. Below the transition
the error bars show the errors in the estimation of $\langle Y_c
\rangle$: the time average of $Y_c$. The solid line is the
bifurcation curve \protect{(\ref{MT420})} close to threshold. The
empty circle at $\Delta=3$ shows the result of the hydrodynamic
simulation by E. Livne. The dashed line is an interpolation
between the solid line and the empty circle. } \label{fig6}
\end{figure}

Noticeable in Fig. 6 is a systematic discrepancy, within the wide
fluctuation-dominated region, between the positions of the maxima
of $P(|Y_c|)$ and the hydrostatic bifurcation curve computed in
Sec. III. We even cannot exclude a change in the character of
bifurcation caused by the fluctuations (apparently without
shifting the bifurcation point). Indeed, the maxima of $P(|Y_c|)$
at $\Delta=1.0$, $1.3$ and $2.0$ appear to lie on a straight line
passing through the theoretical transition point $\Delta_c \simeq
0.5$.  As $\Delta$ increases further, the discrepancy between the
positions of the maxima of $P(|Y_c|)$ and the theoretical
bifurcation curve goes down \cite{accuracy}. Importantly, the
fluctuation-dominated region $0.3 < \Delta < 1.0$ does include the
hydrostatic transition point $\Delta_c \simeq 0.5$.

We should stress that the failure of hydrostatics is observed at
\textit{intermediate} values of the aspect ratio $\Delta$, when
the hydrodynamic parameters $\eta$ and $f$, and the number of
particles $N$, are fixed. In view of Eq. (\ref{restitution}),
while increasing $\Delta$, one increases the inelasticity of
particle collisions $1-r$. That the hydrostatic theory fails at
intermediate values of the inelasticity, and improves at small
enough, or large enough inelasticities, excludes the inelasticity
itself as the reason for the failure.

\subsection{Simulations with different $N$ 
}

We did a series of simulations with different number of particles
$N$ in order to verify the hydrodynamic scaling and investigate
the $N$-dependence of the (relatively weak) fluctuations well
below and well above $\Delta_c$. These additional simulations were
done for $\Delta=0.1$ and three values of $N$: $5 \cdot 10^3$,
$10^4$ and $1.5 \cdot 10^4$, and for $\Delta=3.0$ and $N=4 \cdot
10^4$.

When varying $N$ at fixed $\Delta$, we kept the hydrodynamic
parameters $\eta=11,050$ and $f=0.025$ constant. Therefore, if the
hydrostatic equations provide a correct leading-order theory of
the steady states far below and far above $\Delta_c$, the
time-averaged steady state values of $Y_c$ should become
$N$-independent for large enough $N$. Figure ~7 shows $Y_c$ versus
time for $\Delta=0.1$ at the four different values of $N$.  One
can see that, in all these cases, the average value of $Y_c$ is
close to zero as expected, while fluctuations are relatively
small. Figure ~8 shows the dynamics of  $Y_c(t)$ for $\Delta=3$
and two different values of $N$: $2 \cdot 10^4$ and $4 \cdot
10^4$. Here the symmetry-breaking is evident, as a dense cluster
develops in a corner. With a moderate accuracy determined by the
relatively high level of fluctuations of $Y_c$, the average values
of $Y_c$ at late times are close to each other. Therefore, well
below and well above $\Delta_c$ the hydrodynamic scaling is
obeyed.

\begin{figure}[ht]
\epsfig{file=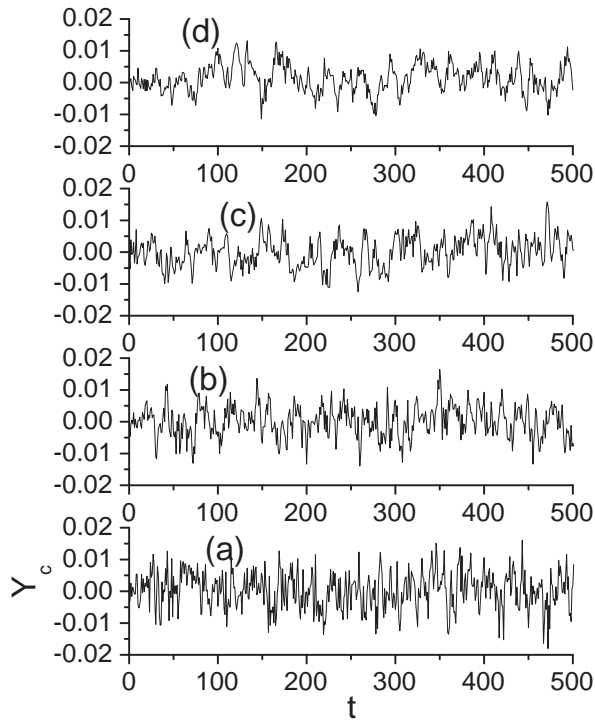, width=8.6cm, clip=}
\caption{$Y_c$ versus time for $\eta=11,050$, $f=0.025$ and
$\Delta=0.1$, for $N=5000$ (a), $10\,000$ (b), $15\,000$ (c) and
$20\,000$ (d), as observed in EMD simulations. Time units are the
same as in Fig. ~4.} \label{fig7}
\end{figure}

\begin{figure}[ht]
\epsfig{file=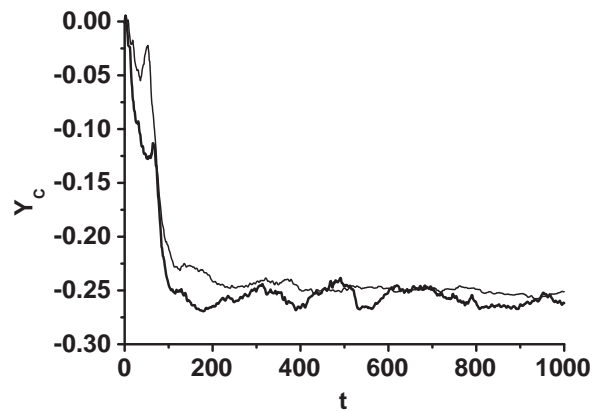, width=8.6cm, clip=} \caption{$Y_c$ versus
time for $\eta=11,050$, $f=0.025$ and $\Delta=3$, for two
different values of $N$, as observed in EMD simulations. The thick
line corresponds to $N=2 \cdot 10^4$, the thin line corresponds to
$N=4 \cdot 10^4$. Time units are the same as in Fig. ~4.}
\label{fig8}
\end{figure}

Simulations with fixed scaled parameters $\eta$, $f$ and $\Delta$,
but different $N$ can also help in identifying the mechanism of
breakdown of the hydrostatic theory at aspect ratios around
$\Delta_c$. Indeed, it is natural to interpret the giant
oscillations, shown in Fig.~4c-e,  in terms of a strong coupling
between the two bifurcated states predicted by the hydrostatic
theory. One possible scenario of this coupling (which we call
Scenario I) relies on the discrete-particle noise, unaccounted for
by granular hydrodynamics. \text {Below} $\Delta_c$, the
discrete-particle noise is expected to cause fluctuations, that is
to broaden the distribution of $Y_c$ as indeed observed in Fig.~5.
If Scenario I is correct, the standard deviation $\sigma$ of $Y_c
(t)$ from its average value  should vanish as $N$ goes to
infinity, at fixed hydrodynamic parameters $\eta$, $f$ and
$\Delta$.

Another possibility (Scenario II) is that the fluctuations persist
in the limit of $N \to \infty$. If this is the case, the
dominating mechanism of fluctuations has a purely hydrodynamic
nature and should be explainable by a \textit{full} hydrodynamic
analysis (as opposed to our hydrostatic analysis, and to the
simplified hydrodynamic simulations that used a model Stokes
friction instead of the full viscosity). Here the coupling between
the two symmetry-broken states  may be due to either an unstable
hydrodynamic mode (Scenario IIa), or a weakly damped mode
(Scenario IIb). In Scenario IIb, $\sigma$ should vanish, as $N \to
\infty$, if one waits for a sufficiently long time. Therefore, to
distinguish between the two sub-scenarios, one should, in addition
to the limit of $N \to \infty$, take the limit of $t \to \infty$.

Obviously, one is unable to take any of these two limits in actual
EMD simulations, where the maximum achievable values of $N$ and
$t$ are limited by the available computer resources. So what was
observed in our EMD simulations with different $N$? Figures~7 and
9 show what happens well below $\Delta_c$, when $N$ increases from
$5\,000$ to $20\,000$. One can see from Fig.~7 that, as $N$ grows,
the high-frequency components of the fluctuations do decrease, but
the low frequency component does not show any pronounced decrease.
Overall, the fluctuation spectrum moves towards the lower
frequencies.  As the result, a good resolution of the
low-frequency part of the power spectrum requires longer and
longer simulations (which rapidly become prohibitively long). This
introduces an additional, non-trivial constraint on simulations
with a large number of particles. A similar situation occurs well
above $\Delta_c$. Figure~8 does indicate that $\sigma$ goes down
as $N$ goes up from $20\,000$ to $40\,000$. However, one also
observes that, as $N$ grows, the role of the low-frequency
components of the fluctuations increases.

Hydrodynamics provides a hint for the mechanism of the ``red
shift" of the power spectrum with an increase of $N$. There are
four hydrodynamic modes in the system: two acoustic modes, the
entropy mode and the shear mode. The frequencies of the acoustic
modes are the highest, as they are determined by the ``ideal"
(non-dissipative) terms in the hydrodynamic equations, and they
scale like the inverse system size. The frequencies of the entropy
and shear modes are much lower, as they are determined by the
transport coefficients: the heat conduction, viscosity and
inelastic loss rate, and  they scale like the inverse
\textit{square} of the system size. In the units of $d=m=T_0=1$,
and at fixed hydrodynamic parameters $\eta$, $f$ and $\Delta$, a
larger $N$ implies a larger system, see Eqs. (\ref{dimensions}).
Correspondingly, as $N$ increases, the characteristic frequencies
of the entropy/shear modes go down much faster than those of the
acoustic modes. Therefore, it seems likely that one of these modes
is responsible for the low-frequency components of the
fluctuations. A related issue is that, in contrast to the
hydrostatic problem (\ref{energy1}), the full time-dependent
hydrodynamic problem has an additional scaled parameter: $d/L_x$.
This parameter describes the role of the dissipative terms
compared to the ``ideal"  terms in the hydrodynamic equations. As
it is clear from Eq. (\ref{dimensions}), when increasing $N$ at
constant $\eta$ and $f$, one reduces this additional parameter.
Therefore, as $N$ increases, the low-frequency shear/entropy modes
should become more and more persistent. As these modes are not
necessarily broad-band, $\sigma$ might cease to provide a good
characterization of the system at large $N$.

Still, if one continues following $\sigma$ as $N$ increases, one
observes (see Fig.~9) that $\sigma$ decreases much slower than the
classic dependence $N^{-1/2}$ characteristic of equilibrium
systems. If one attempts to interpret the decrease of $\sigma$
with an increase of $N$ in terms of an empiric power law, one
obtains an exponent $-0.23$, instead of the classical value of
$-1/2$ for equilibrium systems. Importantly, we did reproduce the
classical $N^{-1/2}$ scaling of $\sigma$ in a control series of
simulations with the same $f$ and $\Delta$, but with $\eta=0$
(elastic collisions). Moreover, a good quantitative agreement was
obtained with a theoretical result for $\sigma$ that directly
follows from the classic expression for the density correlation
function in equilibrium \cite{LL}. We also found that, for the
same total number of particles $N$, the fluctuation levels in the
elastic case are significantly lower than in the inelastic case.
That is, well below $\Delta_c$, the fluctuations, though much
smaller than those observed for $\Delta \sim \Delta_c$, are still
large compared to the elastic case.

\begin{figure}[ht]
\epsfig{file=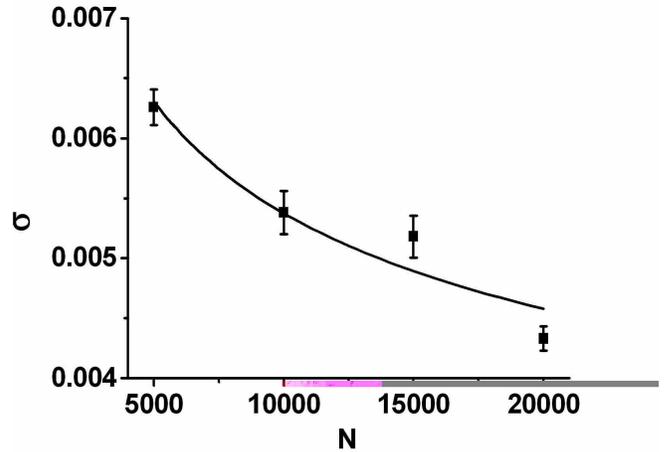, width=8.6cm, clip=} \caption{Shown is
$\sigma$, the standard deviation of $Y_c$ from its (almost zero)
average value, versus the number of particles $N$ for
$\eta=11,050$, $f=0.025$ and $\Delta=0.1$. The symbols show the
simulation results. The curve shows, as a reference, the power-law
dependence $ \sigma =B N^{-\beta}$ with exponent $\beta=-0.23$,
see the text.} \label{fig9}
\end{figure}

To summarize this subsection, our simulations with different $N$
strongly indicate that the hydrostatic equations provide a correct
leading order theory of this system well below and well above
$\Delta_c$. On the other hand, the simulations proved to be
insufficient for determining the mechanism of giant fluctuations
that we observed in this system at $\Delta \sim \Delta_c$. We
cannot even be sure at this point whether the fluctuations (or,
more precisely, their low-frequency components) persist or not as
$N\to\infty$.

\section{Summary and Discussion}
The main results of this work can be summarized in the following
way. Granular hydrostatics, in combination with simplified
hydrodynamic simulations, correctly predict the phase separation
instability in this prototypical driven granular system. Well
above and well below the critical value of the aspect ratio
$\Delta_c$, the hydrostatic theory describes the steady state of
the system well. However, in a wide region of aspect ratios around
$\Delta_c$ the system is dominated by fluctuations, and the
hydrostatic theory fails. The fluctuation levels are anomalously
high even relatively far from the hydrostatic bifurcation point,
and they certainly do not exhibit the classic $N^{-1/2}$ scaling
with the number of particles $N$.

Though we are unable to pinpoint the mechanism of excitation of
the giant fluctuations, we can suggest two different scenarios for
their origin. In Scenario I the fluctuations are driven by
discrete particle noise. Indeed, it is well known that discrete
particle noise can drive relatively large fluctuations in the
vicinity of thresholds of hydrodynamic instabilities
\cite{Hohenberg} and non-equilibrium phase transitions
\cite{Haken}. Fluctuations of this type should vanish as one
increases indefinitely the number of particles in the system,
keeping the hydrodynamic parameters constant.  Unfortunately, our
simulations with different $N$, but fixed $\eta$, $f$ and
$\Delta$,  have been insufficient to prove or disprove this
scenario.

A difficulty with Scenario I is that the fluctuations are so big
in so wide a region of aspect ratios. No anomaly of this type has
been observed in any other symmetry-breaking instability of
granular flow, even with much smaller numbers of particles. As an
example, let us consider for a moment the same system, but
introduce gravity in the $x$ direction. Now the granular gas is
heated from below, and the system exhibits \textit{another}
symmetry-breaking instability: thermal convection, similar to the
Rayleigh-B\'{e}nard convection of classical fluids. The transition
to convection occurs via a supercritical bifurcation
\cite{Ramirez,HMD,KMconvection}. Though EMD simulations of thermal
granular convection \cite{Ramirez} involved only $N=2,300$
particles (which is much less than $N=2 \cdot 10^4$ used in the
present work), a \textit{sharp} supercritical bifurcation was
observed, in agreement with a hydrodynamic analysis
\cite{HMD,KMconvection}. By comparison, the giant fluctuations,
observed in a wide region of $\Delta$ in the present work, are an
anomaly, as one needs some (hydrodynamic?) mechanism of strong
\textit{amplification} of the discrete-particle noise.

If Scenario I proves to be correct, the corresponding theory can
be developed in the framework of Fluctuating Hydrodynamics
\cite{LL}, generalized to granular gases in the limit of nearly
elastic collsions. Fluctuating Hydrodynamics is a Langevin-type
theory that takes into account the discrete character of particles
by adding delta-correlated noise terms in the momentum and energy
equations \cite{LL}. Fluctuating Hydrodynamics is by now well
established for classical fluids in 3D, including non-equilibrium
states \cite{Hohenberg,Mansour}. We should mention here that the
2D case has an additional difficulty. The coupling of fluctuations
here is anomalously strong, even in the elastic case: the
transport coefficients diverge in the thermodynamic limit, except
for a sufficiently dilute gas \cite{2D}. Therefore, one can hope
to generalize the Fluctuating Hydrodynamics to the 2D gas of
inelastic hard spheres in the dilute limit \cite{unfortunately}.
Close to the phase separation threshold, the dilute limit holds
with a reasonable accuracy. It would be interesting to investigate
the phase separation problem in 3D, where important differences in
the fluctuation behavior may occur.

Alternatively, in Scenario II the low-frequency component of the
giant fluctuations has a purely hydrodynamic origin and is driven
either by a presently unknown hydrodynamic instability (Scenario
IIa), or by a long-lived transient mode (Scenario IIb). Effects of
these type are obviously missed by a \textit{hydrostatic}
analysis. They may have also been missed by the time-dependent
hydrodynamic simulation \cite{Livne} that employed a model Stokes
friction, rather than the hard-sphere viscosity, to accelerate the
convergence to a steady state. If Scenario II is correct, the
low-frequency component of the fluctuations should be observable
in hydrodynamic simulations with the true hard-sphere viscosity.
These simulations, therefore, should be an important next step in
the analysis of this fascinating problem.

We are very grateful to E. Livne for doing the hydrodynamic
simulation. We thank A. Barrat, I. Goldhirsch and E. Khain for
useful discussions. This research was supported by Deutsche
Forschungsgemeinschaft (Grant PO 472/5-2), by the Israel Science
Foundation (grant No. 180/02), by the Russian Foundation for Basic
Research (grant No. 02-01-00734) and by Deutscher Akademischer
Austauschdienst.


\begin{references}
\bibitem{accuracy} Notice that, for large aspect ratios,
hydrodynamic simulations are not expected to give a high accuracy
for an entirely different reason. The density of the
symmetry-broken cluster here approaches the close-packing density,
while the accuracy of the interpolating constitutive relations
\cite{Grossman} is not very high.
\end{references}
\end{document}